\documentclass[fleqn,10pt,twocolumn]{wlscirep}
\usepackage{amssymb}
\graphicspath{ {./images/} }
\usepackage{stmaryrd}
\usepackage{amsfonts}
\usepackage{amsmath}
\usepackage{amssymb}
\usepackage{amstext}
\usepackage{accents}
\usepackage{epstopdf}
\usepackage{bm}
\usepackage{xfrac}
\usepackage{units}
\usepackage{braket}
\usepackage{mathtools}
\usepackage[normalem]{ulem}
\usepackage{epsfig}
\usepackage {float}
\usepackage{graphicx}
\usepackage [hang, loose] {subfigure} % to include subfigures
\usepackage{geometry}
\usepackage{filecontents}
\usepackage{hyperref}
\usepackage{color,soul}
\usepackage{xcolor}
\sethlcolor{yellow}

\usepackage{color,soul}

\definecolor{purple}{rgb}{0.5, 0.0, 0.5}

\definecolor{lightblue}{rgb}{.8,.95,1}
\definecolor{LightBlue}{rgb}{.8,.9,1}
\sethlcolor{lightblue}
\definecolor{lightred}{rgb}{1,.8,.8}

\definecolor{LightRed}{rgb}{1,.8,.8}
\definecolor{steelblue}{rgb}{.27,.51,.7}

\sethlcolor{lightred}

\title{Training Data Selection for Accuracy and Transferability of Interatomic Potentials}
\author[1,*]{David Montes de Oca Zapiain}
\author[1]{Mitchell A. Wood}
\author[2]{Nicholas Lubbers}
\author[3]{Carlos Z. Pereyra}
\author[1]{Aidan P. Thompson}
\author[4,*]{Danny Perez}
\affil[1]{Sandia National Laboratories, Albuquerque, NM 87185, USA}
\affil[2]{Computer, Computational, and Statistical Sciences Division, Los Alamos National Laboratory, Los Alamos, NM, 87545 USA}
\affil[3]{Mechanical and Aerospace Engineering, University of California Davis, Davis, California 95616, USA}
\affil[4]{Theoretical Division, Los Alamos National Laboratory, Los Alamos, NM, 87545, USA}

\affil[*]{dmonte@sandia.gov, danny\_perez@lanl.gov}
\keywords{Deep Learning, Information Entropy, Molecular Dynamics, Multi-Scale Modeling}
\begin{abstract}

Advances in machine learning (ML) techniques have enabled the development of interatomic potentials that promise both the accuracy of first principles methods and the low-cost, linear scaling, and parallel efficiency of empirical potentials. 
Despite rapid progress in the last few years, ML-based potentials often struggle to achieve transferability, that is, to provide consistent accuracy across configurations that significantly differ from those used to train the model. 
In order to truly realize the promise of ML-based interatomic potentials, it is therefore imperative to develop systematic and scalable approaches for the generation of diverse training sets that ensure broad coverage of the space of atomic environments. 
This work explores a diverse-by-construction approach that leverages the optimization of the entropy of atomic descriptors to create a very large ($>2\cdot10^{5}$ configurations, $>7\cdot10^{6}$ atomic environments) training set for tungsten in an automated manner, i.e., without any human intervention. 
This dataset is used to train polynomial as well as multiple neural network potentials with different architectures.  
For comparison, a corresponding family of potentials were also trained on an expert-curated dataset for tungsten. 
The models trained to entropy-optimized data exhibited vastly superior transferability compared to the expert-curated models. 
Furthermore, while the models trained with heavy user input (i.e., domain expertise) yield the lowest errors when tested on similar configurations, out-sample predictions are dramatically more robust when the models are trained on a deliberately diverse set of training data.
Herein we demonstrate the development of both accurate and transferable ML potentials using automated and data-driven approaches for generating large and diverse training sets.

\end{abstract}

\begin{document}

\flushbottom
\maketitle
\thispagestyle{empty}
%%%%%%%%%%%%%%%%%%%%%%%%%%%%%%%%%%%%%%%%%%%%%%%%%%%%%%%%%%%%%%
%% INTRODUCTION
%%%%%%%%%%%%%%%%%%%%%%%%%%%%%%%%%%%%%%%%%%%%%%%%%%%%%%%%%%%%%%
\section*{\label{sec:intro}INTRODUCTION}

The rapid adoption of machine learning (ML) methods in virtually all domains of physical science has caused a disruptive shift in the expectations of accuracy versus computational expense of data-driven models. 
The diversity of applications arising from this swell of attention has brought about data-driven models that have accelerated pharmaceutical design\cite{Lounkine_2012_pharma_ref,Ietswaart_pharma_ref_2,Chua_2017_pharma_ref_3}, material design\cite{Panchal_2013,Ramakrishna_2019}, the processing of observations of celestial objects\cite{Davies_2019}, and enabled accurate surrogate models of traditional physical simulations\cite{brunton2016discovering,  montans2019data, patel2021physics}. 
While many of these models have proved extremely powerful, new questions and challenges have arisen due to the uncertainty in model predictions coined as \textit{extrapolations}\cite{Fort_2019}, i.e., when prediction occurs on input that are found outside of the support of the training data. 
Moreover, the accuracy of a machine-learned model can only be quantified using the training itself, or on a subset thereof, held out as validation. 
For that reason, it is often extremely difficult to predict ``real-world" performance where unfamiliar inputs are likely to be encountered.

\begin{figure}[!tb]
	%\centering
	\includegraphics[width=1.0\linewidth]{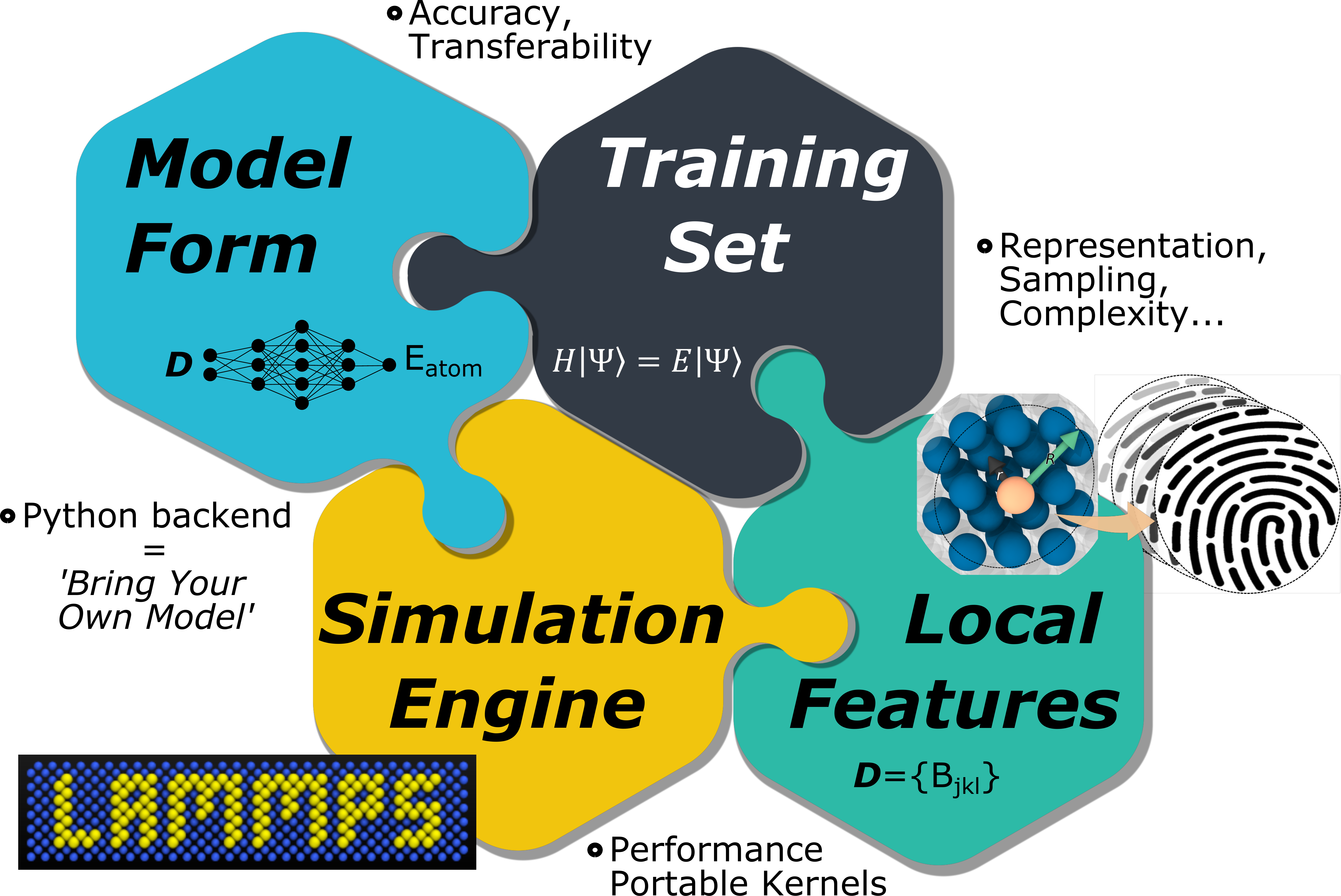}
\caption{ Schematic for the component parts needed for a scalable and accurate Molecular Dynamics simulation utilizing a machine learned interatomic potential. Advances made herein correspond to the \textit{Model Form - Training Set} pair whereas the computational aspects of the remaining pairs have been detailed elsewhere\cite{nguyen2021billion}.}
\label{fig:summary_}

\end{figure}

In this manuscript, we focus on application to classical molecular dynamics (MD), which is a powerful technique for exploring and understanding the behavior of materials at the atomic scale. 
However, performing accurate and robust large-scale MD simulations is not a trivial task because this requires the integration of multiple components, as Figure \ref{fig:summary_} schematically shows.  
A key component is the interatomic potential (IAP)\cite{plimpton2012computational, becker2013considerations, hale2018evaluating, Behler_2007,Behler_2011}, i.e., the model form that maps local atomic environments to energies and forces needed to carry out a finite time integration step. 
An accurate IAP is critical because large-scale MD simulations using traditional quantum \textit{ab initio} calculations such as Density Functional Theory (DFT) are prohibitively expensive beyond a few hundred atoms. Given a local (i.e., short-ranged) IAP, MD simulations can leverage large parallel computers since the calculation of forces can efficiently be decomposed into independent computations that can be concurrently executed \cite{nguyen2021billion}, thus enabling extremely large simulations\cite{zepeda2017probing, germann2008trillion} that would be impossible with direct quantum simulations.
However, a critical limitation of empirical IAPs is that they are approximate models of true physics/chemistry and as such have to be fitted to reproduce reference data from experimental or quantum calculations. 
ML techniques have recently enabled the development of IAPs that are capable of maintaining an accuracy close to that of quantum calculations while retaining evaluation times that scale linearly, presenting significant computational savings over quantum mechanics (QM). 
This is due to their inherent ability to learn complex, non-linear, functional mappings that link the inputs to the desired outputs \cite{Gastegger_2017,Goodfellow_2016,Butler_2018}. 
Nevertheless, despite significant advances in the complexity of the behavior that can be captured with ML-based models, machine learning interatomic potentials (MLIAPs) often struggle to achieve broad transferability \cite{szlachta2014accuracy,Podryabinkin_2017,Smith_2021}. 

Indeed, increasing the complexity of the ML models, while useful to improve accuracy, is not sufficient to achieve transferability. 
In fact, it could even be detrimental, as more data could be required to adequately constrain a more complex model. 
In other words, the more flexible the IAP model form, the more critical the choice of the training data becomes. 
Generating a proper training set is not a trivial task given the fact that the feature space the models need to adequately sample and characterize, which is the space that describes local atomic environments, is extremely high dimensional. 
Consequenlty, MLIAPs are typically trained on a set of configurations that are deemed to be most physically relevant for a given material and/or to a given application area \cite{Zuo2020}, as determined by domain experts. 
Numerous examples now demonstrate that high-accuracy predictions are often achieved for configurations similar to those found in the training set \cite{Bartok_2010_b,Wood_2019,Thompson_2015,Podryabinkin_2017,Smith_2021,Drautz_2019}. 
However, the generic nature of the descriptors used to characterize local environment\cite{Bartok_2013,Bartok_2010_b,Drautz_2019,musil2021physics}, coupled with the inherent challenge in extrapolating with ML methods, often lead to poor transferability. 
This challenge can be addressed in two ways: i) by injecting more physics into the ML architectures so as to constrain predictions or reproduce known limits \cite{patel2021physics, patel2022thermodynamically, mao2020physics}, or ii) by using larger and more diverse training sets to train the MLIAP so that a majority of the atomic environments encountered during MD simulations are found within the support of the training data\cite{bernstein2019novo}.
In this work, we explore the second option, which is generally applicable to all materials and ML architectures.
While conceptually straightforward, a challenge presented to this approach is that relying on domain expertise to guide the generation of very large training sets is not scalable, and runs the risk of introducing anthropogenic biases \cite{Jia_2019}. 
Therefore, the development of scalable, user-agnostic, and data-driven protocols for creating very large and diverse training sets is much preferable.
Additionally it is worth stressing that this problem, and solutions herein, are not confined to the development of IAP, but apply to nearly all supervised ML training problems. 

The key objective of this paper is to demonstrate a new, diverse-by-construction, approach for the curation of training sets for general-purpose ML potentials whose transferability greatly exceeds what can be expected of potentials trained to human-crafted datasets. 
We believe that this approach fills an important niche for the many application domains where it is extremely difficult for users to {\em a priori} identify/enumerate all relevant atomic configuration (e.g., when simulating radiation damage, shock loading, complex microstructure/defect effects, etc.).

In the following, we demonstrate improvements to the training set generation process based on the concept of entropy optimization of the descriptor distribution \cite{Karabin_2020}. 
We leverage this framework to generate a very large ($>2\cdot10^{5}$ configurations, $>7\cdot10^{6}$ atomic environments) and diverse dataset for tungsten (hereby referred to as the {\em entropy maximized}, EM) set in a completely automated manner. 
This dataset was used to train MLIAP models of atomic energy of various complexity, including neural-network-based potentials, as well as linear \cite{Thompson_2015} and quadratic SNAP potentials \cite{Wood_2018_quadratic}. 
The performance of EM-training was compared to equivalent models trained on a human-curated training set used for developing an MLIAP for W/Be (referred to as the {\em domain-expertise, DE} set below) \cite{szlachta2014accuracy, Wood_2019}.

While Tungsten is chosen as the material of interest, none of the results are specific to the choice of material or elemental species.
The results show that EM-trained models are able to consistently and accurately capture the vast training set.
We demonstrate a very favorable accuracy/transferability trade-off with this novel EM training set, where extremely robust transferability can be obtained at the cost of a relatively small accuracy decrease on configurations deemed important by experts. 
In contrast, potentials trained on conventional DE datasets are shown to be mildly more accurate on configurations similar to those contained in the training set, but to potentially suffer catastrophic increase in errors on configurations unlike those found in the training set.
%Furthermore, EM-trained models provide accurate predictions of the DE dataset, demonstrating that a generic and purely automated training set generation procedure can capture the physics that domain experts deem important. 
%In contrast, the MLIAPs trained on the DE set exhibited the lowest errors when validation occurred on randomly held-out configurations from the the DE set, but very large errors when tested on configurations from the EM set. 
Furthermore, the DE-trained MLIAPs exhibited significant sensitivity to the classes (expert-defined groupings) of configurations that were used for training and validation. 
Specifically, large errors were observed when the training and validation sets contained subjectively different manually-labeled classes of configurations instead of a random split. 
These results highlight the perils of relying on physical intuition and manual enumeration to construct training sets. 
In contrast, the EM-based approach is inherently scalable and is fully automatable, since it does not rely on human input. Therefore, making the generation of very large diverse-by-design training sets, and hence the development of accurate and transferable MLIAPs, possible.

%%%%%%%%%%%%%%%%%%%%%%%%%%%%%%%%%%%%%%%%%%%%%%%%%%%%%%%%%%%%%%
%% RESULTS
%%%%%%%%%%%%%%%%%%%%%%%%%%%%%%%%%%%%%%%%%%%%%%%%%%%%%%%%%%%%%%
\section*{\label{sec:results}RESULTS AND DISCUSSION}

\subsection*{\label{sec:results_rep}Representation and Sampling}

In order to avoid over fitting complex ML models, many popular techniques such as compressive sensing, principal components, drop-out testing, etc., are employed to test if the model complexity exceeds that of the the underlying data. 
  
While we only have tested a pair of training sets, the objective is to understand how complex of a model is needed to capture the ground truth (quantum accuracy) embedded in either training set. 
It is important to note that even though the discussion will be focused on the specific application to interatomic potentials, the complexity of the training set is an important factor in any supervised machine learning task.

We first characterize and contrast the characteristics of training sets generated by the two aforementioned methods. 
The domain expertise(DE) constructed dataset, containing about $1\cdot10^4$ configurations and $3\cdot10^5$ atomic environments, was created in order to parameterize a potential for W/Be \cite{Wood_2019}, building on a previous dataset developed for W \cite{szlachta2014accuracy}. 
The data was manually generated and labeled into twelve groups (elastic deformations of BCC crystals, liquids, dislocation cores, vacancies, etc.) so as to cover a range of properties commonly thought to be important \cite{Zuo2020, Shapeev16}. 
The second set was generated using an automated method that aims at optimizing the entropy of the descriptor distribution \cite{Karabin_2020}, as described in Sec.\ref{sec:entropy} \textit{Maximization of the Descriptor Entropy}.

\begin{figure*}[!tb]
\centering
\includegraphics[width=1.0\linewidth]{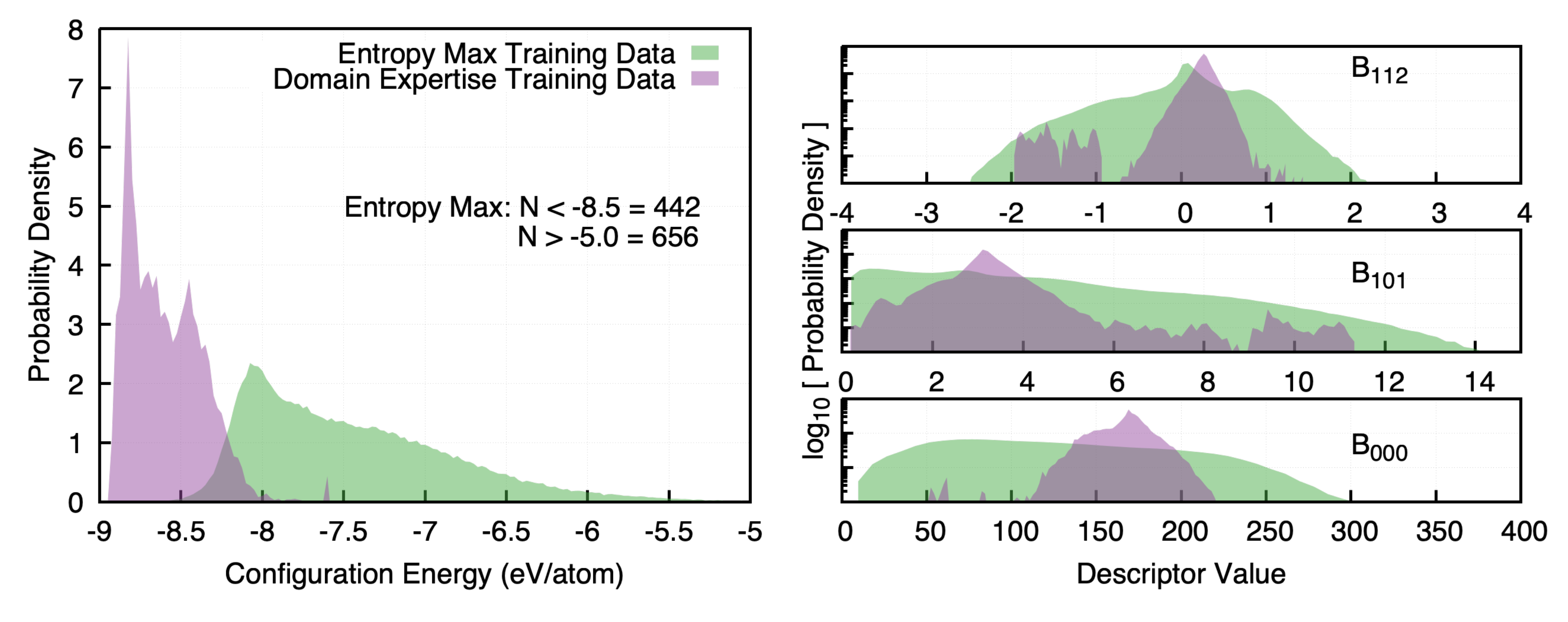}
\caption{Distributions of energy and bispectrum components of the configurations generated using the entropy maximization (EM) framework and using domain expertise (DE). (Left) Distribution of potential energy predicted by density functional theory. (Right) Distribution of three different low-rank bispectrum components, see \textit{Methods} for details on descriptor values. }
\label{fig:compare_set}
\end{figure*}

Figure \ref{fig:compare_set} compares the DE set to the EM set in terms of the distribution of configuration energy (left panel) and of three low-rank bispectrum components\cite{Bartok_2010, Thompson_2015} (right panel), which we use to represent the atomic environment (cf. \textit{Methods})\cite{Bartok_2010_b, Thompson_2015}. 
Observed in the energy distribution, the EM set is very broad in comparison with the DE set which is strongly peaked at energies close to the known ground-state BCC energy of -8.9 eV/atom, extending to about -7.5 eV/atom. 
In contrast, the energy distribution of the EM set spans energies between -8.5 and -5.0 eV/atom, peaking at around -8 eV/atom. 
Due to the large number of configurations in the EM set, a sizeable number of configurations are also located in the tails of the distribution; 442(0.19\%) below -8.5 eV/atom and 656(0.29\%) above -5.0 eV/atom. 
While the overlap between the energy distributions of the two sets is limited, it will be shown below that MLIAPs trained on the EM set can accurately capture the energetics of these low-energy DE configurations. 

The three different probability density plots (Fig. \ref{fig:compare_set} right panel) show that the descriptor distribution of the EM set is also much broader and more uniform than the DE set (note the logarithmic probability scale). 
To quantitatively compare the bispectrum distributions, we compared to covariance of both datasets in the frame of reference where the DE data has mean zero and unit covariance in the space spanned by the first 55 bispectrum components (as defined by a ZCA whitening transformation). 
In that same frame, the EM descriptor distribution is broader in 51 dimension and at least 5x broader in 27 dimensions, strongly suggesting that Fig. \ref{fig:compare_set} provides a representative view of the actual behavior in high dimension. 
In the four dimensions where the entropy set is slightly narrower, it is so by a factor between 0.5 and 0.9.
This is consistent with the relative performance of the MLIAPs trained and tested on the different datasets, as will be shown in Sec.\ref{sec:transferability} \textit{Transferability}.  

In order to further demonstrate the increased diversity and uniformity of the EM dataset relative to that of the DE set, we performed Principal Component Analysis (PCA) in order to represent the training data of each dataset with a reduced number of dimensions \cite{Jolliffe_2005,Suh_2002}. 
\textit{Supplemental Note 1} shows that the PCA representation of the DE set isolates into multiple clusters significantly in the first two PC dimensions. 
On the other hand the PC representation of the EM set is a single cluster that is dense and compact in the first two PC dimensions. 
Therefore, further demonstrating the overall more uniform and diverse sampling of the descriptor space in EM training.
A projection of the DE training into the principal components of EM data is given Supplemental Figure 14 resulting in the same conclusion.

It is important to note that the choice of bispectrum components as descriptors of the local atomic environment is not the only possible representation.
Entropy maximization in another descriptor space (e.g. moment tensors
%\cite{Shapeev16}
, atomic cluster expansion
%\cite{Drautz_2019}
)will result in a somewhat different distribution of configurations. 
However, we expect the qualitative features highlighted here using bispectrum descriptors, including the fact that the DE set is largely contained within the support of the EM set, to be robust. 
\textit{Supplemental Note 13} further demonstrates this as it shows atomic environments generated using the EM method with bispectrum components still encompass DE when using Smooth Overlap of Atomic positions (SOAP) as descriptors. \cite{Bartok_2010,Bartok_2010_b,Bartok_2013,Rosenbrock_2017}

\subsection*{\label{sec:results_acc} Accuracy Limits}

While a very diverse training set is {\em a priori} preferable with regards to transferability, accurately capturing the energetics of such a diverse set of configurations could prove challenging. 
It is therefore important to assess how the choice of model form/complexity affects the relative performance of MLIAPs trained and tested on the EM and DE sets. 
To do so, we consider a broad range of models, ranging from linear, quadratic, and neural network forms, in increasing levels of complexity. 
In all cases, the input features are the bispectrum descriptors that were used to generate and characterize the training sets. 
The models used can be seen as generalizations of the original SNAP approach\cite{Thompson_2015, Wood_2018_quadratic, cusentino2020explicit}. 
Details of the model forms tested, as well as information on the fitting process can be found in the \textit{Methods} section and in the examples provided in the Supplemental Information.

\begin{figure*}[!tb]
	\centering
	\includegraphics[width=1.0\linewidth]{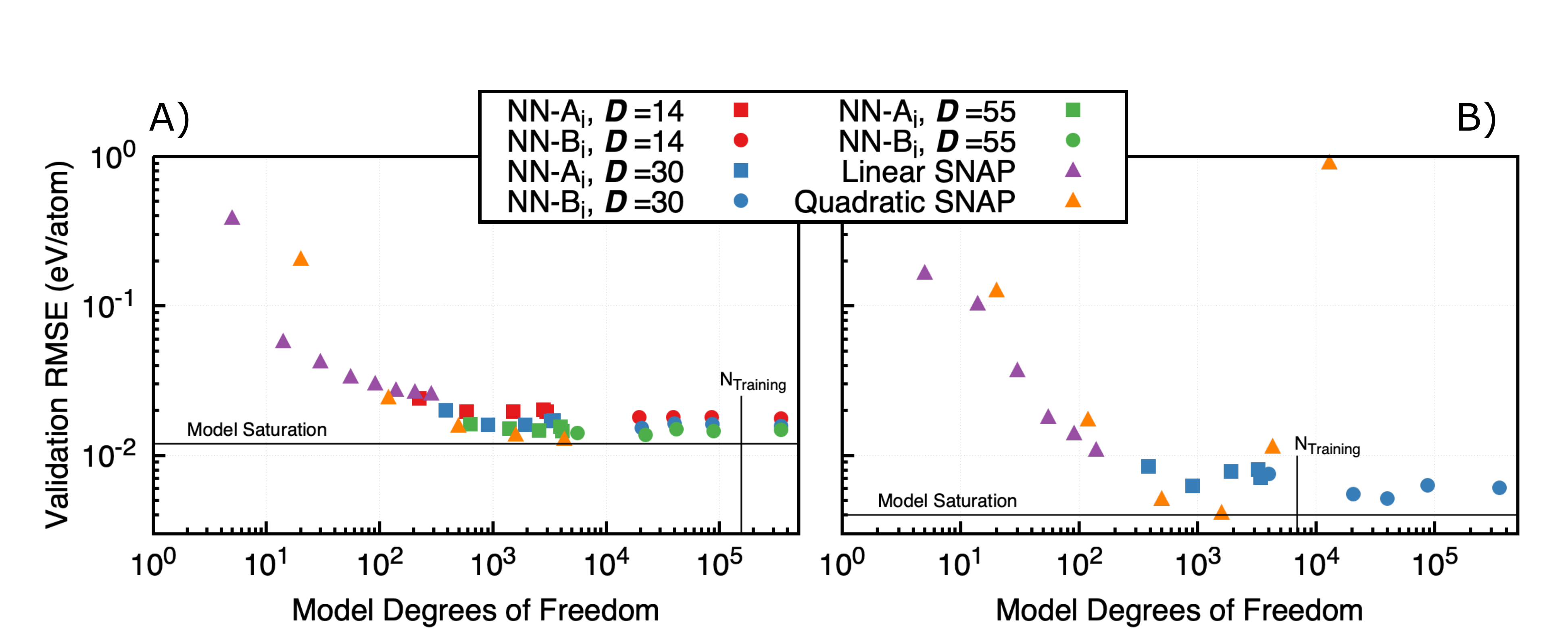}
\caption{ RMSE validation errors for as a function of the number of degrees of freedom for different types of models trained and tested on distinct random samples from their respective dataset. A) EM set; B) DE set. Each model form and complexity shows a saturation of the validation errors indicated by the solid horizontal line. Slight improvement in NN predictions is seen with increasing $\textbf{\textit{D}}$.} 
\label{fig:dof_vs_rmse}
\end{figure*}

The accuracy of the trained models was assessed by quantifying the error on a validation set of configurations randomly held-out of the training process. 
We first evaluated the performance of each one of the different models for predicting the energy of configurations that were generated with the matching framework (i.e., EM or DE) on which the model was trained. 
Figure \ref{fig:dof_vs_rmse} reports the accuracy, quantified as the Root Mean Squared Error (RMSE), of the models on their respective validation set.

In order to compare models of variable complexity, Figure \ref{fig:dof_vs_rmse} displays validation RMSE against the number of free parameters $N_{DoF}$ that are optimized in the regression step of the different ML models.
For the simplified ML models (linear and quadratic SNAP) $N_{DoF}$ is determined uniquely by the number of descriptors, $\textbf{\textit{D}}=\{B_{jkl}\}$, where $\textbf{\textit{D}}$ denotes the number of bispectrum component used to characterized the atomic environments and is determined by the level of truncation used. 
However, for NN models $N_{DoF}$ also depends on the number of hidden layers and the number of nodes per layer (cf. \textit{Methods} section for more details).
Figure \ref{fig:dof_vs_rmse} A) and B) show that in spite of the different nature of the models evaluated, the performance of all model classes remarkably asymptotes to roughly the same error, including the deep NNs where $N_{DoF} \ge N_{Training}$. 
This asymptote occurs after about $10^3$ DoF, which is much less than the training size of $\sim 10^5$ for the EM set. 
The value of the limiting error is observed to slightly decrease with increasing number of descriptors.
Finally, the errors are observed to saturate at a lower value when training and validating on the DE set ($\sim 4\cdot 10^{-3}$eV/atom) than on the EM set ($\sim 10^{-2}$eV/atom). 
This is perhaps unsurprising given the comparatively more compact and less diverse nature of the DE set, which makes it more likely that the validation set contains configurations that are relatively similar to configurations in the training set, a point that we will expand on the following paragraphs.

Figure \ref{fig:dof_vs_rmse} B) also shows that NN models surprisingly do not overfit to the training data even when $N_{DoF} \ge N_{Training}$, where one could expect the RMSE value of the validation set to increase. 
This is presumably caused by the non-convex nature loss function which makes it more difficult to access very low loss minima that would lead to overfitting. 
In addition, it is important to mention that our training protocol also reduced the learning rate when the validation loss plateaued (cf. \textit{Methods} for the detailed protocol). 
In contrast, quadratic SNAP models, cf. Figure \ref{fig:dof_vs_rmse} B) --- for which the loss function is convex --- show clear signs of overfitting to the DE set when $N_{DoF} \simeq N_{Training}$ , an observation that was previously reported \cite{Wood_2018_quadratic, Zuo2020}. 

These results demonstrate that the details of the MLIAP's architectures appear to have limited impact on the ultimate accuracy when training and validation data are sampled from similar distributions, so long as the model is sufficiently flexible. 
The following section shows that the choice of the training data, not the model form, is the key factor that determines the transferability of the models. 

\subsection*{\label{sec:transferability} Transferability}

While the accuracy/transferability trade-off has been evident for many years for traditional IAPs that rely on simple functional forms, the development of general ML approaches with very large numbers of DoF in principle opens the door to MLIAPs that would be both accurate and transferable. 
However, as discussed above, the introduction of flexible and generic model forms can in fact be expected to make the selection of the training set one of the most critical factors in the development of robust MLIAPs. 

\begin{figure*}[!tb]
	\centering
	\includegraphics[width=1\linewidth]{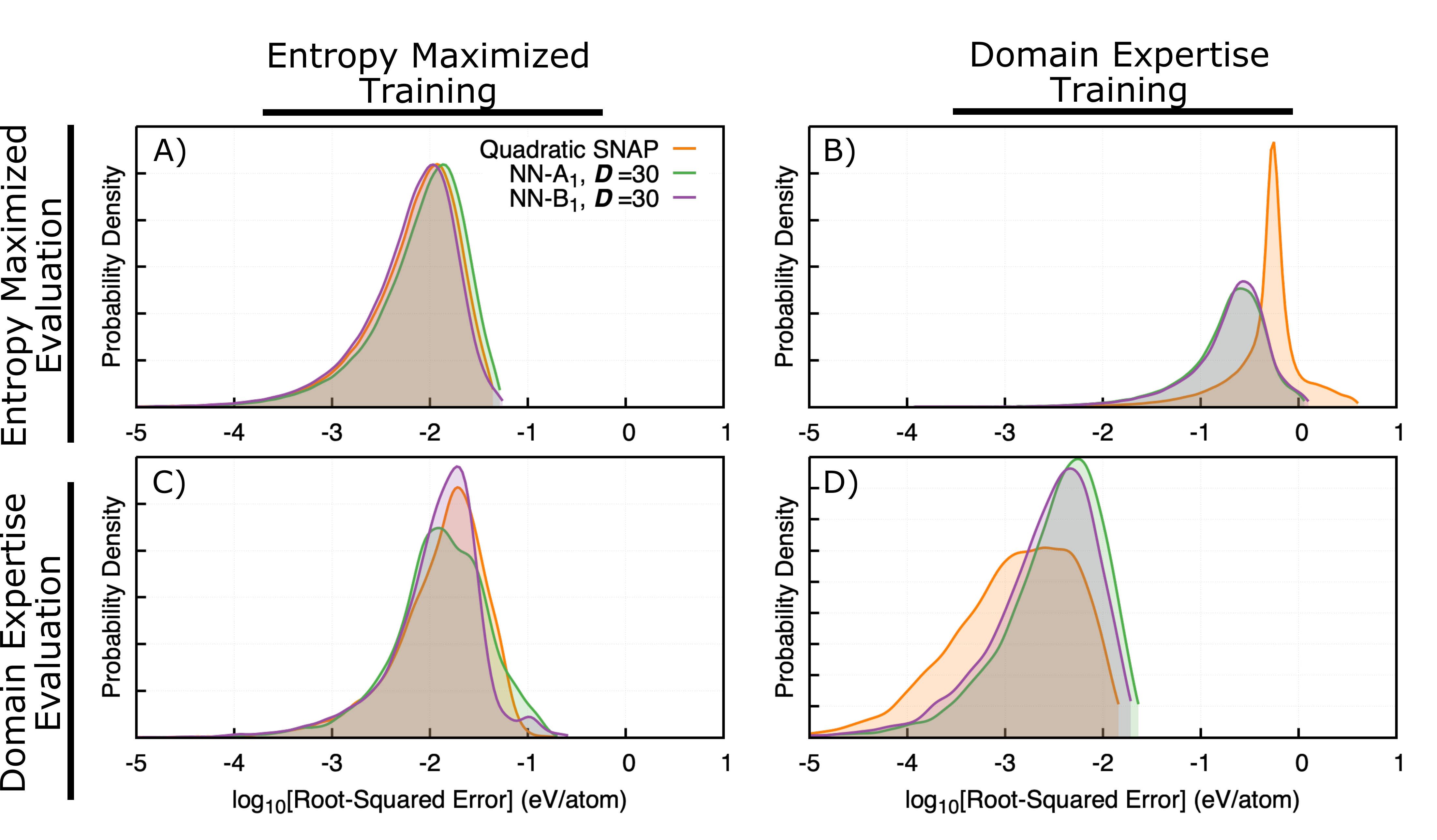}
\caption{Distribution of RSE errors for different combination of training and testing data. A) Trained on EM and validated on EM; B) trained on DE and validated on EM; C) Trained on EM and validated on DE; D) trained on DE and validated on DE. Only errors within three standard deviations from the observed mean value , about $\sim 99\%$ of the data, are reported to clearly convey the shape of the distribution. }
\label{fig:quadplot}
\end{figure*}

In order to assess the relative transferability of the models trained on the different datasets, we select three models for further analysis: the NN-$A_{1}$, NN-$B_{1}$ and a quadratic SNAP model, all using an angular momentum limit of $J_{\mathrm{max}}=3$ which corresponds to $\textbf{\textit{D}}=30$.
These three models have 383, 3965 and 495 degrees of freedom, respectively. 
Also, all of these models are on the saturation regime where the model performance asymptotes to roughly the same error. 
Figure \ref{fig:quadplot} reports the distribution of the root squared error (RSE) from the different models trained and validated on the four possible combinations of DE and EM. 
Specifically, Figure \ref{fig:quadplot} A) and D) report the RSE distributions predicted for configurations selected from the same set as the one used to train the model (e.g., when both training and validation are done on DE data), while panels B) and C) reports the error distributions when validation configurations are chosen from a different set (e.g., when training is done on DE data but validation is done on EM data and vice versa). 
The detailed errors corresponding to these four different possibilities, as well as to additional MLIAP models, are reported in the Supplemental Notes 4, 5, 6, and 7.

Figure \ref{fig:quadplot} A) and D) show that the both sets of models exhibit low errors when predicting the energy of randomly held-out configurations sampled from the set the model was trained on (DE or EM), consistent with the results shown in the previous section. 
In both cases, the distribution peaks around $10^{-2}$ eV/atom and rapidly decays for larger errors, with long tails toward smaller error values. 
The models trained and validated on the DE data (panel D)) show a slightly heavier tail at low errors than the model trained and validated on the EM data, in agreement with the slightly lower asymptotic errors reported in Fig.\ref{fig:dof_vs_rmse}. 
Otherwise, the behavior of the different models is similar, except for the quadratic-SNAP model from panel D) which shows slightly lower errors. 

Transferability of a model is quantified by predicting on configurations sampled from a different dataset than the one used for training, Figure \ref{fig:quadplot} B) and C). 
The performance of the two sets of models now show dramatic differences. 
Figure \ref{fig:quadplot} B) shows a very large increase in errors, by almost two orders magnitude, when predicting the energy of configurations sampled from the EM set using models trained to DE data.
Supplemental Note 9 addresses how these prediction errors are concentrated with respect to the high energy configurations that are clear extrapolations of the model.

In contrast, Figure \ref{fig:quadplot} C) shows only a modest increase in error when predicting the energy of configurations sampled from the DE set using models trained to EM data. 
In other words, models trained on a compact dataset that is concentrated in a small region of descriptor space (such as the DE set) can be very accurate, but only for predictions that are similar to the training data because excursions that force the trained MLAIPs to extrapolate out of the support of their training data leads to extremely large errors. 
Conversely, models trained to a very broad and diverse dataset might have comparatively slightly larger errors when validating over the same diverse dataset, but, perhaps unsurprisingly, show very good and consistent performance when tested on a dataset that is contained within the support of its training data, where testing points can be readily accessed by interpolation. 
This numerical experiment clearly demonstrates that transferability of a given model is critically influenced by the choice of data that is used to trained the model and that very large and diverse datasets ensure both high accuracy and high transferability.

The presented results also suggests that the fact that the distribution of energies of the EM set decays very quickly as it approaches the BCC ground state (cf., Fig.\ \ref{fig:compare_set}), did not affect the performance of the models when testing on configurations from the DE set, whose distribution is strongly peaked at low energies, \textit{see also Supplemental Note 9}.

In order to determine whether the large errors observed when DE-trained models are validated on EM data are attributed to high-energy configurations, that could be argued to be irrelevant in most conditions of practical interest, different partitions of the DE set into training and validation data were also investigated. 
Therefore, instead of partitioning using a random split of the data, the training/validation partitions were instead guided by the manual labeling of the DE set into distinct configuration groups. 
In this case, {\em entire groups} were either assigned to the training {\em or} to the validation set in a random fashion, so that configurations from one group can only be found in either the training or the validation set, but not in both. 
This way, we limit (but not rigorously exclude) the possibility that very similar configurations are found in both the training and validation sets. 
The validation errors measured with this new scheme dramatically increase by one to two orders of magnitude for quadratic-SNAP potentials, as compared to the random hold-out approach. 
Therefore, clearly showing that even well-behaved, expert-selected, configurations can be poorly captured by MLIAPs when no similar configurations are present in the training set. 
Consequently, further demonstrating that MLIAP should not be used to extrapolate to new classes of configurations, even if these configurations are not dramatically different from those found in the training set (e.g., different classes of defects in the same crystalline environment). 
A more detailed analysis can be found in \textit{Supplemental Note 8}.
As a result, the results shown in this work instead suggest that in order to ensure robust transferability one requires the generation of very large and diverse training sets that fully encompass the physically-relevant region for applications, so that extrapolation is never, or at least very rarely, required. 

\begin{figure}[!tb]
%\centering
\includegraphics[width=1.0\linewidth]{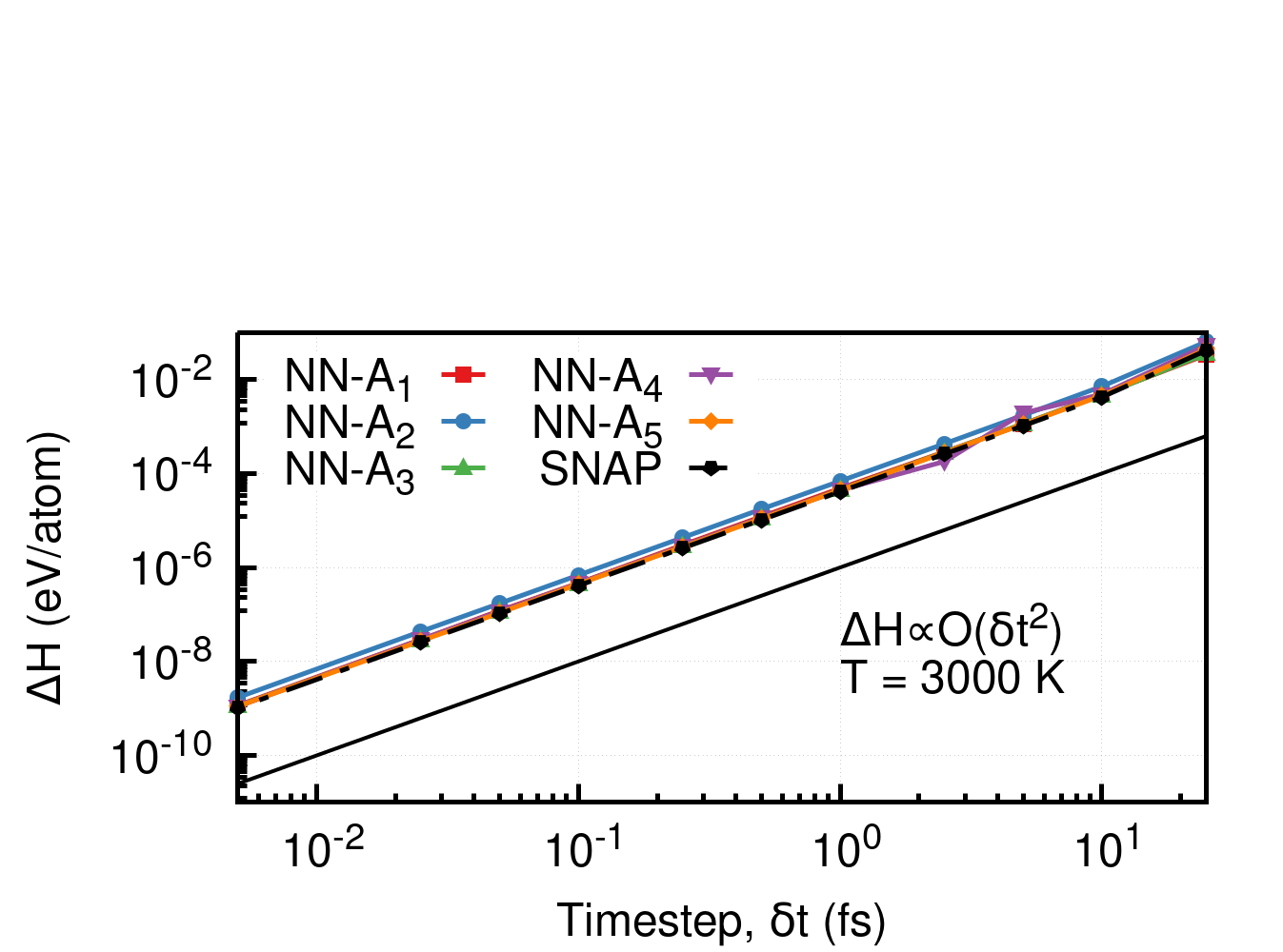}

\caption{Energy conservation as a function of MD timestep size for the Type A NN models and the linear SNAP model with EM training. In all cases, a tungsten BCC supercell was simulated under NVE dynamics at 3000~K for 7.5 ps. The energy deviation was calculated by equation \ref{eqn:fluctuation_relative_diff}. All the models exhibit asymptotic second order accuracy in $\delta t$, characteristic of the St{\"o}rmer-Verlet time discretization.  Higher order deviations  emerge only at $\delta t > 10$~fs, close to the stability limit. This demonstrates that the NN models yield energy and force predictions that are consistent, smooth, and bounded.}
\label{fig:stability}
\end{figure}

Finally, we demonstrate that the resultant NN models are numerically and physical stable by testing them in production MD simulations using the LAMMPS code\cite{LAMMPSweb, Thompson_2022}.
Figure \ref{fig:stability} shows the deviation in energy as a function of MD timestep $\delta t$ for the Type A NN models trained using the EM data set.
In all cases, the degree of energy conservation is comparable to that of the SNAP linear model and exhibits asymptotic second order accuracy in the timestep, as expected for the St{\"o}rmer-Verlet time discretization used by LAMMPS. Higher order deviations emerge only at $\delta t > 10$~fs, close to the stability limit determined by the curvature of the underlying potential energy surface of tungsten. 
All of the models were integrated into the LAMMPS software suite using the recently developed \emph{ML-IAP} package, completing the link between \textit{Model Form} and \textit{Simulation Engine} in Figure \ref{fig:summary_}. 
The \emph{ML-IAP} package enables the integration of arbitrary neural network potentials into the atomistic software suite of LAMMPS, regardless of the training method used. 
As a result, all the models developed in this work can be used to perform simulations with the accuracy of quantum methods (i.e. direct transcription of the energy surface defined by suitably diverse DFT training data).
To conclude, coupling this code package  with the universal training set generation outlined in this work enables a seamless integration of ML models into LAMMPS and thus will enable a breadth of research hitherto unmatched.

\section*{\label{sec:disc}Conclusion}
The present work demonstrates a needed change in the characterization of ML models that is motivated by the goal of transferable interatomic potentials thereby avoiding known pitfalls of extrapolation.
Counter-intuitive results presented here showed that model accuracy saturates even when the model flexibility increases and thereby re-directs attention to what is included as training data when assessing the overall quality of an interatomic potential.
Machine learned interatomic potentials differ from traditional empirical potentials (simple functional forms derived from physics/chemistry of bonding) in this assessment of the trained space wherein the accuracy of an empirical potential is quantified on the ability to capture domain expertise selected materials properties.
Transferring these practices to the training of MLIAP demonstrated that a physically motivated, user expertise, approach for defining training configurations fails to yield MLIAP capable of having desired transferability. 
Nevertheless, this work also demonstrated that transferability can be achieved by producing a training set that maximizes the volume of descriptor space such that the model rarely extrapolates, where high errors are expected, even when this results in high-energy, far from equilibrium states of the material. 
Initial efforts to generate MLIAP from hybridized training sets is promising, see Supplemental Note 11.
In addition, and as a point that is important for the community of molecular dynamics users, the entropy optimized training sets used to generate the transferable MLIAPs  are descriptor agnostic, material independent, and automated with little to no user input tuning. 
Also, novel software advances in LAMMPS now allow for any ML model form to be used as an interatomic potential in a MD simulation.
This is an important scientific advancement because it allows for subsequent research that utilizes these highly accurate and transferable ML models to be used within a code package that is actively utilized by tens of thousands of researchers.

Beyond the use case of IAP, the protocol presented in this work for training set generation is something that many data-sparse ML applications can take advantage of when characterizing the accuracy of a generated model. Therefore, it should be expected that ML practitioners report regression errors, but also now to quantify the complexity of the training data in order for end users to understand where to expect interpolation versus extrapolations in a more quantitative fashion.

%%%%%%%%%%%%%%%%%%%%%%%%%%%%%%%%%%%%%%%%%%%%%%%%%%%%%%%%%%%%%%
%% METHODS
%%%%%%%%%%%%%%%%%%%%%%%%%%%%%%%%%%%%%%%%%%%%%%%%%%%%%%%%%%%%%%

%Danny reviewed this section too.
\section*{METHODS}\label{sec:methods}
\subsection*{\label{sec:entropy}Maximization of the Descriptor Entropy}

The data-driven EM set was generated using an entropy-optimization approach introduced in previous work \cite{Karabin_2020}. This framework aims at generating a training set that 
is i) diverse, so as to cover a space of configurations that encompasses most configurations that could potentially be observed in actual MD simulation, and ii) non-redundant, in order to avoid spending computational resources characterizing many instances of the same local atomic environments. To do so, we introduce the so-called {\em descriptor entropy} as an objective function that can be systematically optimized. In what follows, the local environment around each atom $i$ is described by a vector of descriptors $\textbf{\textit D}_i$ of length $m$. These descriptors can be arbitrary differentiable functions of the atomic positions around the target atom. 
In this work, the $\textbf{\textit D}_i$ are taken to be the bispectrum components which were introduced in the development of the GAP potentials \cite{Bartok_2010,Bartok_2010_b,Bartok_2013}, and then adopted in the SNAP approach \cite{Thompson_2015, Wood_2018}. To avoid excessive roughness on the entropy surface in high dimension, we used the five lowest-order bispectrum components ($m=5$) in the optimization procedure. 
As reported above, we nonetheless observe very significant broadening of the descriptor distribution compared to the DE set in almost all directions of the 55-dimensional space induced by the lowest-order bispectrum components. This behavior, which will be studied in detail in an upcoming publication, was also observed in the original publication \cite{Karabin_2020}.

The diversity of local environments within a given configuration of the system can be quantified by the entropy of the $m-$dimensional descriptor distribution $S(\textbf{\textit{\{D\}}})$. High entropy reflects a high diversity of atomic environments within a given configuration, while low entropy corresponds to high similarity between atomic environments. The descriptor entropy is therefore an ideal objective function in order to create a diverse dataset. The creation of high entropy structures can be equivalently recast as the sampling of low-energy configurations on an effective potential energy surface given by minus the descriptor entropy. This enables the training set generation procedure to be implemented in the same molecular dynamics code that will be used to carry out the simulations.

The effective potential is of the form:

\begin{eqnarray} \label{eq:effective_energy}
    V_\mathrm{entropy} &=& V_\mathrm{repulsive} -K S(\{\textbf{\textit D}\})
    \end{eqnarray}
    
where $V_\mathrm{repulsive}$ is a simple pairwise repulsive potential that mimics a hard-core exclusion volume, thereby prohibiting close approach between atoms, and $S(\{\textbf{\textit D}\})$ is a non-parametric estimator of the descriptor entropy based on first-neighbor distances in descriptor space \cite{Beirlant_1997}. The addition of  $V_\mathrm{repulsive}$ is essential to avoid generating nonphysical configurations that would prevent convergence of the DFT calculations. As a result, this effective potential can be used to carry out either molecular-dynamics-based annealing or direct minimization, as discussed in the original publication \cite{Karabin_2020}. 

\begin{figure}[!tb]
%\centering
\includegraphics[width=1.0\linewidth]{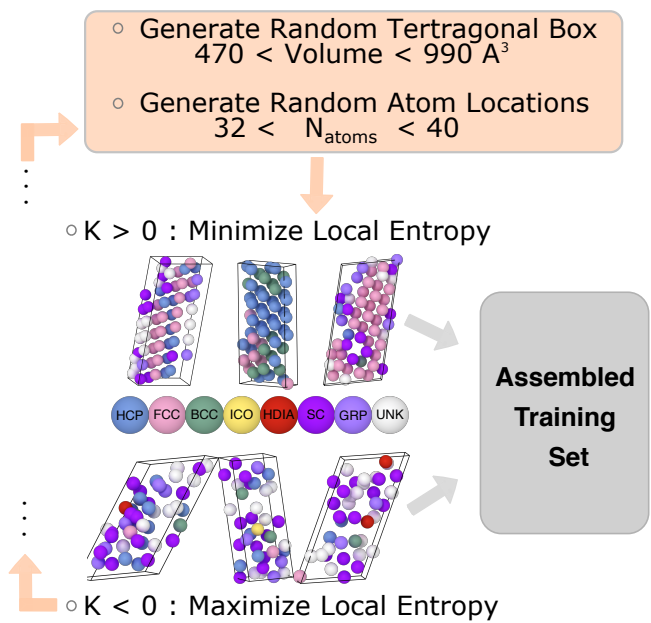}
\caption{Automated workflow that generates both entropy maximized and minimized atomic structures given Equation \ref{eq:effective_energy}. Representative structures of either type are colored by local crystal structure. Entropy minimized structures are mostly crystalline but with defects, while entropy maximized structures are largely amorphous and/or show non-closed packed structure types.}
\label{fig:entflow}
\end{figure}

A possible limitation of this approach is that regions of descriptor space corresponding to crystalline configurations, which are key to many materials science applications, might be under-sampled, as entropy maximization promotes configurations where each atomic environment differs from others. In order to avoid this possibility, the dataset also contains "entropy-minimized" configurations, which are obtained using the same procedure as the entropy-maximized ones, except that that sign of $K$ in Eq.\ \ref{eq:effective_energy} is reversed, leading to configurations where order is promoted instead of suppressed. It is important to acknowledge that the type of local order (e.g., FCC or BCC) that is promoted through entropy minimization is not pre-specified by this approach, the data generation procedure is captured in Figure \ref{fig:entflow}.

%\begin{enumerate}
%    \item Generate a random tetragonal box with volumes ranging between 470 and 990 $\AA^3$
%    \item Insert a random number of atoms (between 32 and 40) at random locations
%    \item Set $K$ to a positive value
%    \item Locally minimize  $V_\mathrm{entropy}$  
%    \item Save the configuration
%    \item Set $K$ to a negative value
%    \item Locally maximize  $V_\mathrm{entropy}$  
%    \item Save the configuration
%    \item Return to 1.
%\end{enumerate}

This loop was repeated $N/2$ times, generating a total of $N=223,660$ configurations, half of which are entropy-minimized, and half entropy-maximized.
The range of atom counts and box volumes is chosen with respect to ambient density of Tungsten ($15.9~\AA^3/atom$).
Note that except for very small systems, this sampling procedure typically does not converge to the global minimum of $V_\mathrm{entropy}$, but instead remains trapped in local minima of the effective potential energy surface. Different initial random starting points (w.r.t. to initial atomic positions and cell sizes and shapes) will converge to different final states. As shown in Fig. \ref{fig:compare_set}, this allows for the generation of a wide range of different configurations, instead of repeatedly generating  structures that are internally diverse but very similar to each other. The difficulty of converging to the global entropy optimum is therefore a positive feature in this case.   
The same effect also occurs in the case of entropy minimization, where this procedure yields crystals that contain stacking defects, grain boundaries, line and point defects, etc., with perfect crystal being only rarely generated. 

The energy and forces acting on the atoms were then obtained with the VASP DFT code using the GGA exchange correlation functional with an energy cutoff was set to 600 eV, and a 2x2x2  Monkhorst-Pack k-point grid used.\cite{kresse1993ab,kresse1996efficient,blochl1994projector,kresse1999ultrasoft}
The calculations were converged to an SCF energy threshold of $10^{-8}$ eV. 
Imposing a constant plane wave energy cutoff and k-point spacing for all of the diverse configurations is certainly an approximation, and was done to automate the generation of these training labels. 
Since the entropy maximization method does not need these DFT results to generate new structures, stricter DFT settings can be applied \textit{a posteriori}.

The main advantage of this method in contrast with conventional approaches that rely on domain knowledge is that it is fully automated and executed at scale because no human intervention is required. In other words, the generated training set was not curated {\em  a posteriori } to manually prune or add configurations, and the weight of the different configurations in the regression was not adjusted. The training configurations are further material-independent, except for the choice of the exclusion radius of the repulsive potential and the range of densities that was explored. As a result, this means that in the case of pure materials, configurations generated using generic parameter values can simply be rescaled based on the known ground-state density of the target material. Therefore, the method naturally lends itself to high throughput data generation, as every training configuration can be generated and characterized with DFT in parallel, up to some computational resource limit.

Note that entropy optimization differs in philosophy from some recently proposed active learning approaches \cite{Podryabinkin_2017,Vandermause_2020,Smith_2021} where training sets are iteratively enriched by using a previous generation of the MLIAP to generate new candidate configuration using MD simulations. Candidates are added to the training set whenever a measure of the uncertainty of the prediction reaches a threshold value. These methods are appealing as they also allow for the automation of the training set curation process, and because they generate configurations that can be argued to be thermodynamically relevant. Nevertheless, the diversity of the training sets generated active learning approaches ultimately relies on the efficiency of MD as a sampler of diverse configurations. However, many potential energy surfaces are extremely rough, which can make them very difficult to systematically explore using methods based on naturally evolving MD trajectories. Take for example a configuration of atoms that has a high energy barrier to a new state of interest. Unless a high temperature is set in MD, which skews the thermodynamically relevant states, these rare events will dictate the rate new training is added. Consequently,  in active learning methods the selection of the initial configuration from which MD simulations are launched then becomes very important. In contrast, entropy optimization explicitly biases the dynamics so as to cover as much of the feature space as possible using an artificial effective energy landscape that maximizes the amount of diversity contained in each configuration. Note however that both approaches can easily be combined by substituting entropy-optimization for MD as the sampler used within an iterative loop. Finally, notice that some active learning approaches, specifically those based on the d-optimality criterion of Shapeev and collaborators \cite{Podryabinkin_2017}, also build on the insight that extrapolation should be avoided in order to identify candidate configurations that should be added to the training set.

\subsection*{Neural Network Models}

Neural networks are highly flexible models capable of accurately estimating the underlying function that connects a set of inputs to its corresponding output values from available observations \cite{Lecun_2015,Goodfellow_2016}. Feed Forward Neural Networks (FFNN) are quite versatile  models that can be tailored to predict the results for a wide variety of applications and fields \cite{Feng_2019,Wang_2018,GAO_2020}. FFNNs have been successfully used to develop IAPs \cite{Behler_2007,Behler_2011,Sosso_2012,Tang_2020}. In this work, we train different FFNNs to learn the mapping between the local atomic environment (characterized using the bispectrum components \cite{Bartok_2010,Bartok_2010_b,Bartok_2013,Thompson_2015}) and the resultant energy of each atom (denoted as $E_i$). The total energy of the system is subsequently obtained by summing the atomic energies  (i.e., $E_{tot}=\sum_i E_i$). Note that the atomic energies are not available from DFT, so training only considers total energies of entire configurations. 
The different neural networks are trained by minimizing the squared error that quantifies the discrepancy between the values predicted by the model (i.e., the FFNN) and a set of ``ground-truth" output values obtained for the training set. In this work we trained two different sets of FFNNs, the EM and DE sets described above, using the same training protocol for both.

The training protocol starts by selecting a subset of the training set on which to train the model. In this work we used 70\% (randomly sampled) of the configurations to train the model. In addition, we used 10\% (again randomly sampled) as a validation set and the remaining 20\% as the test set. The validation set and the test set are not used directly for the training on the model. These sets are used to monitor for over-fitting and to critically assess the ability of the trained model to generalize to new/unseen data. Partitioning a given data set into training, validation, and testing is a common strategy in deep learning model development.

\begin{figure}[!tb]
%\centering
\includegraphics[width=1.0\linewidth]{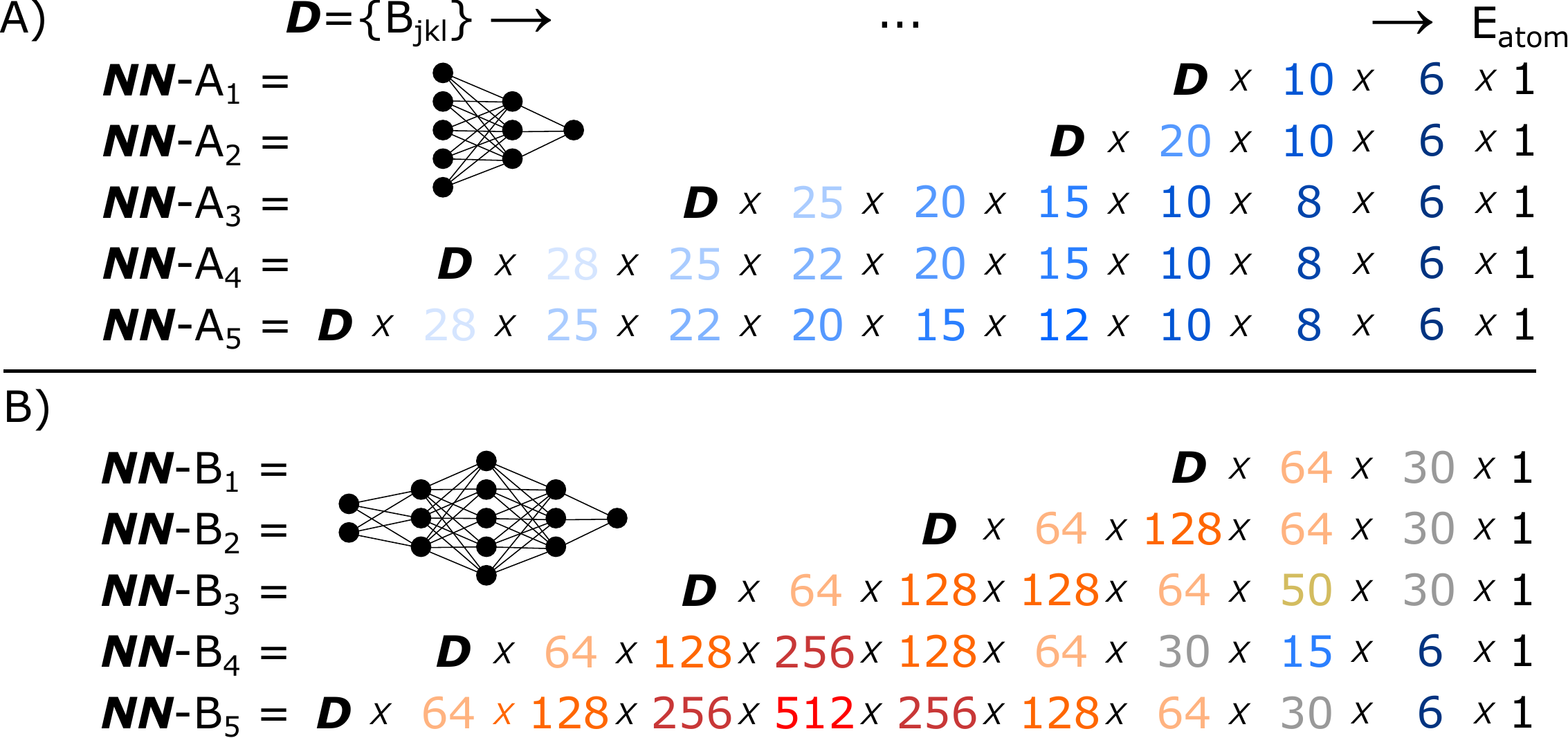}
\caption{Visual representation of the two neural network architecture types: A) type A step-down and B) type B expand-and-contract. The input is a vector of $\textbf{\textit D}$ descriptors of the atomic environment of one atom.  These are passed through each layer in the network to yield the atomic energy $E_i$ of the atom. Integer factors indicate the number of nodes in each layer. The total number of nodes defines the number of degree of freedom $N_{DoF}$ for each model. Recall from Fig. \ref{fig:dof_vs_rmse}
that $\textbf{\textit D}$ is 14, 30, or 55. 
Notice that $\textbf{\textit D}$
only affects the nodes in the input layer since the nodes on the other layers remain unchanged.
}
\label{fig:NNArch}
\end{figure}

This work considered ten different neural networks architectures. Five of the architectures systematically reduced the number of features before predicting the energy in the last layer. These neural networks are denoted as Type A (step-down) and their architecture is illustrated in Figure \ref{fig:NNArch} (A). The activation function used between each layer is the SoftPlus activation function and is applied to all the layers (after transforming the inputs adequately) except the final output layer because it is the one that predicts the energy associated to the input bispectrum components. The other five  neural networks initially increased the number of features and subsequently decreased the number of features before predicting the energy in the last layer and are denoted Type B(expand-then-contract).  Similar to Type A networks, the activation function used is the SoftPlus function.  Figure \ref{fig:NNArch} (B) illustrates the architectures of the Type B neural networks. \textit{Supplemental Note 2} details the procedure leveraged for selecting the optimal learning rate and batch size and \textit{Supplemental Note 3} details the procedure used for selecting the optimal featurization of the bispectrum components for training deep-learning potentials.

Each one of the ten different neural networks was trained for 800 epochs using the optimal values for the learning rate and batch-size previously identified. Furthermore, we used the ADAM optimizer \cite{Kingma_2014_adam} and incorporated a learning rate scheduler that reduced the learning rate by half if the validation loss did not change by $1\cdot10^{-4}$ over 50 epochs. After each network was trained we assessed its accuracy by comparing its energy prediction to the ground-truth (obtained with \textit{ab-initio} calculations) using the Root Squared Error (RSE). 

\subsection*{Simulation Stability}

We tested the numerical and physical stability of all of the models by running realistic molecular dynamics simulations in the LAMMPS atomistic simulation code\cite{LAMMPSweb, Thompson_2022}. The most important characteristic of any classical potential is the degree to which it conserves energy when used to model Hamiltonian or NVE dynamics. Theoretically, under these conditions, the total energy or Hamiltonian $H = T + V$ is a constant of the motion, while the kinetic energy $T$ and potential energy $V$ fluctuate equally and oppositely. In practice, the extent to which the total energy is conserved is strongly affected by the timestep size $\delta t$, as well as the time discretization scheme, and any pathologies in the potential energy surface and corresponding forces. Because LAMMPS uses
 St{\"o}rmer-Verlet time discretization that is both time reversible and symplectic\cite{Hairer_2006}, a well-behaved potential should exhibit no
energy drift and the small random variations in energy that do occur should have a mean amplitude that is second order in the timestep size.   
We characterize both of these effects by simulating a fixed-length trajectory with a range of different timestep sizes and sample the change in total energy relative to the initial state.  The average energy deviaton is defined to be 
\begin{align}
    \begin{split}
    \Delta H(\delta t) &=\dfrac{1}{nN}{\sum_{i=1}^{n}\lvert H(t_i; \delta t) - H(t_0)\rvert} \, ,
    \end{split}
    \label{eqn:fluctuation_relative_diff}
\end{align}
where $H(t_i; \delta t)$ is the total energy sampled at time $t_i$ from a trajectory with timestep $\delta t$, $n$ is the total number of samples, and $N$ is the number of atoms. 

All the NVE simulations were initialized with 16 tungsten atoms in a BCC lattice with periodic boundary conditions, equilibrated at a temperature of 3000~K, close to the melting point.  Each simulation was run for a total simulation time of 7.5~ps and the number of time samples $n$ was 1000.
All calculations were performed using the publicly released version of LAMMPS from November 2021. In addition to the base code, LAMMPS was compiled with the \emph{ML-IAP}, \emph{PYTHON}, and \emph{ML-SNAP} packages. The \emph{MLIAP\_ENABLE\_PYTHON} and \emph{BUILD\_SHARED\_LIBS} compile flags were set. 
An example LAMMPS input script has been included in the Supplemental Information.

The dependence of energy deviation on timestep size for some representative models is shown in Figure \ref{fig:stability}. In all cases, we observe asymptotic second order accuracy, as expected for the St{\"o}rmer-Verlet time discretization used by LAMMPS. Higher order deviations emerge only at $\delta t > 10$~fs, close to the stability limit determined by the curvature of the underlying potential energy surface of tungsten.

\newcommand{\br}{{\bf r}}
\newcommand{\bu}{{\bf u}}
\newcommand{\bU}{{\bf U}}
\newcommand{\bB}{{\bf B}}
\newcommand{\balpha}{{\boldsymbol\alpha}}
\newcommand{\bbeta}{{\boldsymbol\beta}}

\subsection*{Bispectrum Components and SNAP Potentials}

The entropy maximization effective potential, the neural network potentials, and the SNAP potentials described in this paper all use the bispectrum components as descriptors of the local environment of each atom. These were originally proposed by Bartok \emph{et al.} \cite{Bartok_2010,Bartok_2010_b,Bartok_2013} and then adopted in the SNAP approach \cite{Thompson_2015, Wood_2018}.

In the linear SNAP potential, the atomic energy of an atom $i$ is expressed as a sum of the bispectrum components ${\bf B}_i$ for that atom, while for quadratic SNAP, the pairwise products of these descriptors are also included, weighted by regression coefficients 
\begin{eqnarray}
E_i({\bf r}^{N}) & = & 
\bbeta \cdot {\bB}_i + 
\frac{1}{2} {\bB}_i \cdot \balpha \cdot {\bB}_i \,,
\label{snapE}
\end{eqnarray}
where the symmetric matrix $\balpha$ and the vector $\bbeta$ are constant linear coefficients whose values are determined in training. The bispectrum components are real, rotationally invariant triple-products of  four-dimensional hyperspherical harmonics $\bU_j$ \cite{Bartok_2010}
    \begin{eqnarray}
        \label{eqn:z}
        B_{j_1j_2j}  &=& \bU_{j_1} \otimes_{j_1j_2}^j \bU_{j_2} \colon \bU_j^* \,, \end{eqnarray}
where symbol $\otimes_{j_1j_2}^j$ indicates a Clebsch-Gordan product of two matrices of arbitrary rank, while $:$ corresponds to an element-wise scalar product of two matrices of equal rank. For structures containing atoms of a single chemical element, the $\bU_j$ are defined to be  
    \begin{eqnarray}
        \label{eqn:u}
        \bU_j &=& \sum_{r_{ik} < R}~{f_c(r_{ik}) \bu_j(\br_{ik})} \,,
    \end{eqnarray}
where the summation is over all neighbor atoms $k$ within the cutoff distance $R$.  The radial cutoff function $f_c(r)$ ensures that atomic contributions go smoothly to zero as $r$ approaches $R$ from below. The hyperspherical harmonics $\bu_j$ are also known as Wigner U-matrices, each of rank $2j+1$, and the index $j$ can take half-integer values $\{0, \frac{1}{2},1,\frac{3}{2},\ldots\}$.
They form a complete orthogonal basis for functions defined on $S_3$, the unit sphere in four dimensions.\cite{Varshalovich88,Bartok_2013} The relative position of each neighbor atom ${\br}_{ik}= (x,y,z)$ is mapped to a point on $S_3$ defined by the three polar angles $\psi$, $\theta$, and $\varphi$ according to the transformation $\psi = \pi r/r_0$, $\cos\theta = z/r$, and $\tan\varphi = x/y$. The bispectrum components defined in this way have been shown to form a particular subset of third rank invariants arising from the atomic cluster expansion \cite{Lysogorskiy2021}. The vector of descriptors ${\bB}_i$ for atom $i$ introduced in Eq.~\ref{snapE} is a flattened list of elements $B_{j_1j_2j}$ restricted to $0 \le j_2 \le j_1 \le j \le J$, so that the number of unique bispectrum components scales as $\mathcal{O}(J^3)$. In the current work, $J$ values of 1, 2, 3, and 4, are used, yielding descriptor vectors ${\bB}_i$ of length $\textbf{\textit D}$ = 5, 14, 30, and 55, respectively.  The radial cutoff value used for entropy maximization, neural network and SNAP models was $R =4.73$~\AA~.

%%%%%%%%%%%%%%%%%%%%%%%%%%%%%%%%%%%%%%%%%%%%%%%%%%%%%%%%%%%%%%%%%%%%%%%%%%%%%%%%
\section*{Acknowledgements}
%%%%%%%%%%%%%%%%%%%%%%%%%%%%%%%%%%%%%%%%%%%%%%%%%%%%%%%%%%%%%%%%%%%%%%%%%%%%%%%%

%All authors thank \hl{XXXX} for their detailed review and edits. 
%M.A.W,  A.P.T and D.P. acknowledge support from the Exascale Computing Project (17-SC-20-SC), a collaborative effort of the U.S. Department of Energy Office of Science and the National Nuclear Security Administration.
%D.M.Z., N.L, M.A.W., C.Z.P, and A.P.T. acknowledge support by the U.S. Department of Energy, Office of Fusion Energy Sciences (OFES) under Field Work Proposal Number 20-023149.
%Sandia National Laboratories is a multimission laboratory managed and operated by National Technology \& Engineering Solutions of Sandia, LLC, a wholly owned subsidiary of Honeywell International Inc., for the U.S. Department of Energy’s National Nuclear Security Administration under contract DE-NA0003525.
%Los Alamos National Laboratory is operated by Triad National Security LLC, for the National Nuclear Security administration of the U.S. DOE under Contract No. 89233218CNA0000001.

%alternative acknowledgements

The development of the entropy maximization method and the generation of the training data was supported by the Exascale Computing Project (17-SC-20-SC), a collaborative effort of the U.S. Department of Energy Office of Science and the National Nuclear Security Administration. The training of the various MLIAP models and the comparative performance analysis was supported by the U.S. Department of Energy, Office of Fusion Energy Sciences (OFES) under Field Work Proposal Number 20-023149. Sandia National Laboratories is a multimission laboratory managed and operated by National Technology \& Engineering Solutions of Sandia, LLC, a wholly owned subsidiary of Honeywell International Inc., for the U.S. Department of Energy’s National Nuclear Security Administration under contract DE-NA0003525.
Los Alamos National Laboratory is operated by Triad National Security LLC, for the National Nuclear Security administration of the U.S. DOE under Contract No. 89233218CNA0000001.

\section*{AUTHOR CONTRIBUTIONS}
D.M.Z., N.L. and M.A.W. implemented fitting code, fit, and evaluated NN and SNAP models.
D.P. and M.A.W. generated training data.
C.Z.P., N.L. and A.P.T. implemented and evaluated models in LAMMPS.
All authors participated in conceiving the research, discussing the results, and writing of the manuscript.

\section*{COMPETING INTERESTS}
The authors declare no competing interests.

\section*{ADDITIONAL INFORMATION}
{\bf Supplementary information} is available for this paper. \\

\noindent {\bf Correspondence} and requests for materials should be addressed to D.M.Z.(dmonte@sandia.gov) and D.P.(danny\_perez@lanl.gov).

%%%%%%%%%%%%%%%%%%%%%%%%%%%%%%%%%%%%%%%%%%%%%

% \bibliographystyle{elsarticle-num}
% \bibliographystyle{elsarticle-harv}
\bibliography{main_npj}
\clearpage
%%%%%%%%%%%%%%%%%%%%%%%%%%%%%%%%%%%%%%%%%%%%%

\end{document}